\documentclass[]{JHEP3} 

\JHEPspecialurl{http://jhep.sissa.it/JOURNAL/JHEP3.tar.gz}

\usepackage{epsfig,multicol,bbm}
\usepackage{epstopdf}

\newcommand\fverb{\setbox\fverbbox=\hbox\bgroup\verb}
\newcommand\fverbdo{\egroup\medskip\noindent%
			\fbox{\unhbox\fverbbox}\ }
\newcommand\fverbit{\egroup\item[\fbox{\unhbox\fverbbox}]}
\newbox\fverbbox

\title{Natural Neutrino Dark Energy}

\author{Ilya Gurwich\\
	Physics Department, Ben-Gurion University, Beer-Sheva 84105, Israel\\
	E-mail: \email{gurwichphys@gmail.com}}

\abstract{A new class of neutrino dark energy models is presented.
	The new models are characterized by the lack of exotic particles
	or couplings that violate the standard model symmetry.
	It is shown that these models lead to several concrete predictions
	for the dark energy equation of state, as well as possible effects
	on the cosmic structure formation.
	These predictions, can be verified (or disproved) with future experiments.
	At this point, the strongest constraints on these models are obtained from
	big bang nucleosynthesis, and lead to new bounds on the mass of the
	lightest neutrino.}


\begin{document}

\section{Introduction}


It has been over a decade since observations of supernovae revealed that
the expansion of the universe seems to be accelerating \cite{DE_discovery}. 
Today in addition to supernova observations, the acceleration of the universe,
is supported by data from WMAP \cite{WMAP}, SDSS \cite{SDSS} and more
\cite{Hubble_DEC}. The source of this acceleration, the so called "dark energy"
(DE), remains a mystery, since it cannot be any of the well known cosmological
components.
The only previously known energy source that can play the role of DE is the
vacuum (or zero-point) energy.
The magnitude of the zero-point energy density depends on the QFT cutoff.
However, any natural cutoff, generates an enormous vacuum energy, up to 120
orders of magnitude greater than the value needed to account for the observed
acceleration of the universe.
Thus, in addition to the strange nature of DE, one must account for its "tiny"
value ($\rho_{\rm DE}\sim \left(10^{-3}eV\right)^4$), the same value responsible
for the "cosmic coincidence" - the fact that today, DE is of the same
order of magnitude as the cold dark matter density, despite the fact that the
ratio between them changes as the cosmic scale factor cubed.

Countless theories have been presented to explain the source of cosmic
acceleration and its value.
Most of them include either modified gravity \cite{DGP,ModGrav}
or exotic sources of energy \cite{Quintessence, Chaplygin}, deviating far from
the standard model (SM).
Since strong deviations from known physics are usually accompanied by
numerous degrees of freedom, testing for these theories is very difficult,
in spite of the advancing observational constraints on DE
\cite{WMAP, SDSS, Hubble_DEC, DEC}.
In addition, most theories seem to be unable to naturally generate the correct
value for DE, usually leaving it as another degree of freedom,
resulting in the need for fine-tuning.
Finding ourselves in such a vast landscape of alternative theories, emphasizes
the need for a theory with minimal deviations from known physics and preferably
no fine tuning.

\medskip
neutrino masses are the closest known fundamental energy scale to
the scale of DE \cite{RPP,CNB}.
This implies that neutrinos might be connected to DE,
and that a theory with an underlying link between them
would have less additional degrees of freedom.


This idea is not new, and various methods to link neutrino physics to DE,
have been investigated
\cite{NeutMix_DE, CLEP_DE, GoldstoneB_DE, Vacuum_DE,NDE1,NDE2}.
One of the leading models for neutrino DE is mass varying neutrinos
(MAVANs) \cite{MaVaNs_basic, MaVaNs}.
This class of models includes coupling between neutrinos and a scalar field
(the Acceleron).
The DE "tracks" the matter density to a critical point, when the temperature
of the cosmic background and the mass of the neutrino are comparable,
and then becomes constant and generates the acceleration we witness today.
All the models, mentioned above, have specific issues, but one issue is common
to all of them:
It is unnatural for the neutrino to be the only source of DE.
For example, in the case of MAVANs, there is no reason for the Acceleron to
couple only to the neutrino.
Being a neutral scalar particle, coupling to a neutrino must imply
coupling to the electron (and by the same token, the muon and the tau).
A natural model of neutrino DE cannot include such an exotic
coupling that distinguishes the neutrino from the other leptons, thus,
breaking the SM symmetry.

\section{Constructing the Model}

In order to construct a natural neutrino DE (NNDE) model, it is
essential to find a quality that separates the neutrino from the other
leptons. The obvious would be the lack of electric charge.
Models, in which the DE component is coupled to Majoranna mass terms,
naturally account for the neutrino being the only source of DE.
However, since Majoranna masses are typically very high, we lose the
correlation between the mass and the DE scale and thus,
the entire motivation behind neutrino DE.
Therefore, it is necessary to find a different quality that would distinguish
the neutrino.

The three neutrino species, appear to be the lightest fermions.
Particularly, the lightest neutrino is the only known massive particle that {\it may}
still be relativistic in the cosmic background.
The heavier neutrinos are not relativistic
(based on neutrino oscillations \cite{RPP}),
but they are much closer than any other known fermion.
Thus, we can base a NNDE model on the idea that relativistic particles contribute
to DE, while non-relativistic ones do not.
We construct a model, such that the energy density is
\begin{equation}
	\displaystyle{\rho_{\rm DE}\left(a\right)=
	\sum_{\rm i=particles}U\left(\xi_{\rm i}\left(a\right),\chi_{\rm i}\left(a\right),...\right)}~.
	\label{basic}
\end{equation}
Here $\xi_{\rm i}(a),\chi_{\rm i}(a),...$ are scalar functions of certain
parameters of the cosmic ensemble (referring to the particle species $i$).
$\xi_{\rm i}(a)$ and $\chi_{\rm i}(a)$ are almost constant in the ultra-relativistic
regime and diverge as the kinetic energy of the particles drops to zero
(alternatively, as the cosmic scale factor $a\rightarrow\infty$).
The function $U$ must decay to zero as $\xi(a),\chi(a),...$ diverge,
so there is no contribution to DE from non-relativistic particles.
Also, to correlate DE to particle masses, $U\sim m^4$ in the ultra-relativistic
regime, where $m$ is the particle mass.
In eq.~(\ref{basic}), it was not specified, whether we sum over
all elementary particles, fermions or only leptons.
This is to be determined in a specific model and is irrelevant at this point.
Eq.~(\ref{basic}) is not restricted to a specific model and guaranties the following:
\begin{itemize}
	\item No exotic couplings that violate particle physics symmetries are
	necessary, since DE couples evenly to every particle.
	\item	At any given time, DE is dominated by the heaviest
	relativistic particles. Heavier, non relativistic particles, have no
	contribution. Therefore,
	\begin{itemize}
		\item Today, DE is dominated by the (lightest?) neutrino.
		\item	In the past, heavier particles contributed as well and the
		DE density was greater. This, will be referred to as
		primordial DE.
         \end{itemize}
\end{itemize}
The form of eq.~(\ref{basic}) is therefore sufficient to explain why the
neutrino is the only one to contribute to DE today.

\medskip
To proceed, it is necessary to find all the possible parameters $\xi(a),\chi(a)$,...
that satisfy the above mentioned conditions.
The mean energy $E$ must be a factor in all the parameters, since it
defines the relativistic and non-relativistic regimes.
$E$ scales with expansion oppositely to what we have defined;
it is constant in the non relativistic regime and diverges in the
ultra-relativistic one.
Therefore the energy must be suppressed by a quantity that constantly grows
with expansion.
Thus, two simple parameters come to mind, $\xi=El$ and $\chi=E/T$.
Where $l$ is the mean distance between the particles and $T$ is the temperature
of the cosmic background
(T is not necessarily the CMB temperature. For example,
for neutrinos $\displaystyle{T=T_{\rm CMB}\left(4/11\right)^{1/3}}$)

\section{NNDE Phenomenology}

At this point, we can find an expression for the DE equation of state (EOS),
\begin{equation}
	\displaystyle{1+w_{\rm DE}=-\frac{1}{3}\frac{a}{\rho_{\rm DE}}
	\frac{\partial\rho_{\rm DE}}{\partial a}}~.
	\label{EOS_basic}
\end{equation}
Assuming, only a single neutrino species gives a significant contribution,
substituting eq.~(\ref{basic}) into eq.~(\ref{EOS_basic}), we have
\begin{equation}
	\displaystyle{1+w_{\rm DE}=-\frac{1}{3}\frac{a}{U}\left(
	\frac{\partial U}{\partial\xi}\frac{\partial\xi}{\partial a}+
	\frac{\partial U}{\partial\chi}\frac{\partial\chi}{\partial a}\right)}~.
	\label{EOS_gen}
\end{equation}
For cosmic neutrinos \cite{CNB,NDE2}, we can write
\begin{eqnarray}
	\displaystyle{E}&=&\displaystyle{\frac{\displaystyle{\int\frac{4\pi p^2dp}{\left(2\pi\right)^3}
	\frac{\sqrt{m^2+p^2}}{1+\exp(p/T)}}}
	{\displaystyle{\int\frac{4\pi p^2dp}{\left(2\pi\right)^3}
	\frac{1}{1+\exp(p/T)}}}}\,~,
\label{E_mean}
\\
	\displaystyle{l}&=&\displaystyle{\frac{1}
	{\left(\displaystyle{2\int\frac{4\pi p^2dp}{\left(2\pi\right)^3}
	\frac{1}{1+\exp(p/T)}}\right)^{1/3}}}\,~.
\label{l_mean}
\end{eqnarray}
Expanding the square root in eq.~(\ref{E_mean}) via
$\sqrt{m^2+p^2}\simeq\displaystyle{p+m^2/2p}$ and substituting it
along with eq.~(\ref{l_mean}) into eq.~(\ref{EOS_gen}), we have
\begin{equation}
	\displaystyle{w_{\rm DE}\simeq-1-\left(
	\beta_{\xi}\frac{\partial \ln U}{\partial\xi}+
	\beta_{\chi}\frac{\partial \ln U}{\partial\chi}\right)
	\left(\frac{m}{T}\right)^2}~,
	\label{EOS_exp}
\end{equation}
where,
$$
\beta_{\xi}=\displaystyle{\frac{1}{27}\left(\frac{\pi^8}{12\zeta(3)^4}\right)^{1/3}}
\simeq 0.268\,
\qquad
\hbox{and}
\qquad
\beta_{\chi}=\displaystyle{\frac{\pi^2}{54\zeta(3)}}
\simeq 0.152\,
$$
are numerical coefficients.
$U$ and its derivatives are functions of $\xi$ and $\chi$, both of which are constant
to first order, in the ultra-relativistic regime.
Therefore, in the ultra-relativistic regime $T\sim a^{-1}$ is the fastest varying
quantity in eq.~(\ref{EOS_exp}), it follows that to first order,
$1+w_{\rm DE}\sim a^2$.
Expanding eq.~(\ref{EOS_exp}) via powers of $1-a$ we have,
\begin{equation}
	\displaystyle{w_{\rm DE}=w_{0}+w_{\rm a}(1-a)+w_2(1-a)^2}~,
	\label{EOS_phenexp}
\end{equation}
with
\begin{equation}
	\displaystyle{w_{\rm a}=-2w_2=-2(1+w_0)}~,
	\label{EOS_phenexp_rel}
\end{equation}
and
\begin{equation}
	\displaystyle{w_0=-1-\left(
	\beta_{\xi}\frac{\partial \ln U}{\partial\xi}+
	\beta_{\chi}\frac{\partial \ln U}{\partial\chi}\right)
	\left(\frac{m}{T_0}\right)^2}~.
	\label{EOS_phenexp_0}
\end{equation}
Where $T_0$ is the temperature at the present time.
Eq.~(\ref{EOS_phenexp}) is valid even if $w_{\rm DE}$ deviates strongly from
-1, as long as the ratio $m/T$ is small, and thus the particles are relativistic.

Verifying the relation $w_{\rm a}(w_0)$ is a good, parameter independent,
method to test a variety of DE models.
Table~\ref{wa_w0_relations} shows the linear evolution approximation  of the
DE EOS, for different models.
Some of these models are plotted in figure~\ref{wa_w0}.
One can clearly see the distinctive behavior of the different models.

\FIGURE{\epsfig{file=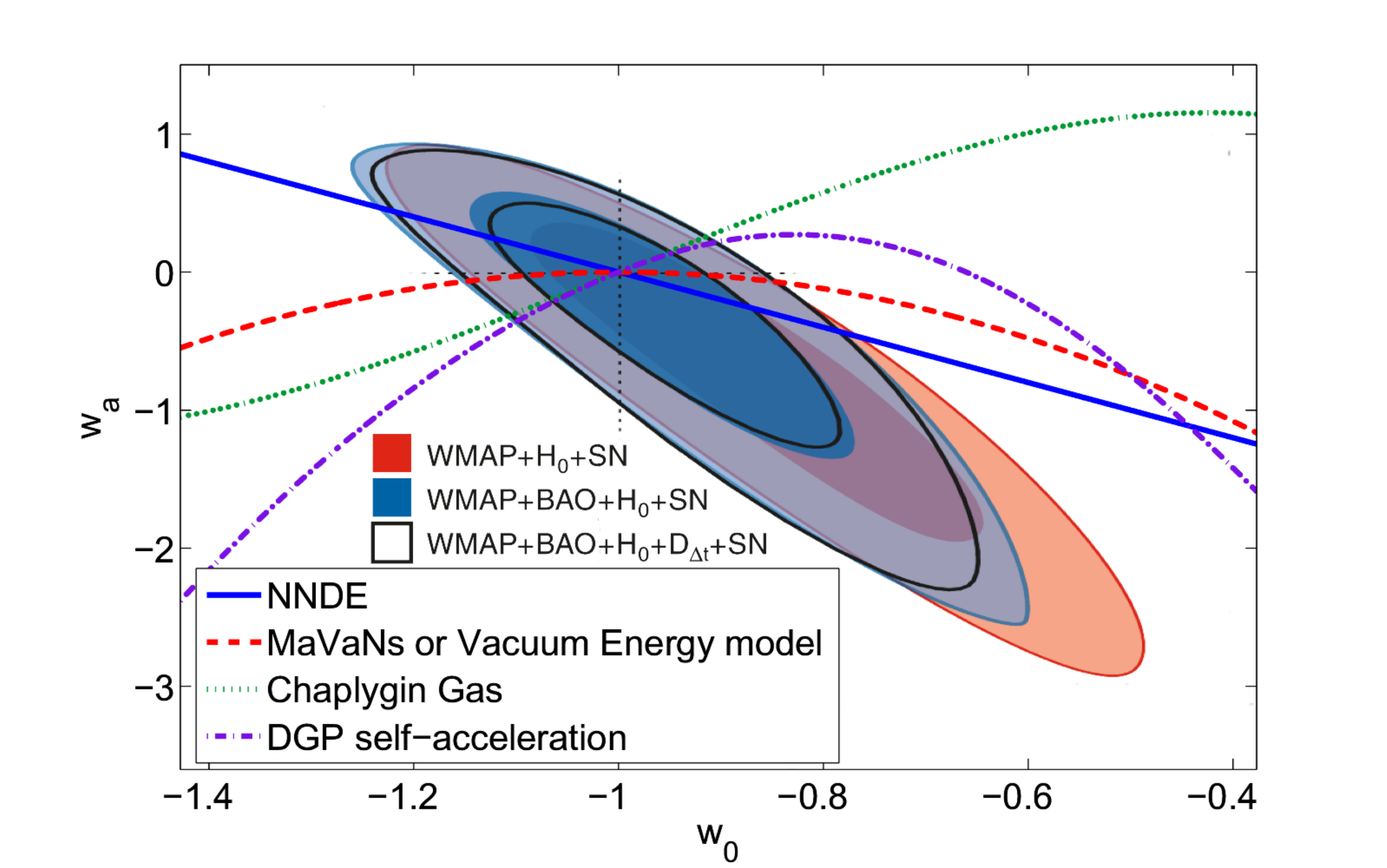,width=15cm} 
        \caption{The linear evolution approximation of the DE EOS, for different
        models.
        The plotted contours signify the observational constraints, taken from
        \cite{WMAP}.
        Within the observational constraints, the different models are clearly
        distinguishable, aside from the DGP and Chaplygin gas curves.}
	\label{wa_w0}}

\TABULAR[t]
{|l|l|}{\hline 
\bf{Model} & ${\bf w_{\rm a}(w_0)}$\\
\hline
NNDE & $w_{\rm a}=-2(1+w_0)$\\
\hline
basic MaVaNs & $w_{\rm a}=-3(1+w_0)^2$\\
\hline
Vacuum fluctuations & $w_{\rm a}=-3(1+w_0)^2$\\
\hline
Chaplygin gas & $w_{\rm a}=3(1+w_0)\left(1-(1+w_0)^2\right)$\\
\hline
DGP & $w_{\rm a}=3(1+w_0)\left(1-2(1+w_0)\left(1+\sqrt{-w_0(1+w_0)}\right)\right)$\\
\hline 
Quintessence Thawing & $-3(1+w_0)\leq w_{\rm a}\leq-(1+w_0)$\\
\hline 
Quintessence Freezing & $-0.2w_0(1+w_0)\lesssim w_{\rm a}\leq-3w_0(1+w_0)$\\
\hline}{The linear evolution approximation of the DE EOS, for different models.
From top to bottom: NNDE, a simple MaVaN model
\cite{MaVaNs_basic}, Vaccum energy model \cite{Vacuum_DE},
Chaplygin gas \cite{Chaplygin}, DGP \cite{DGP},
quintessence thawing and quintessence freezing
\cite{Quintessence}. \label{wa_w0_relations}}

As shown, NNDE corresponds to the thawing range of parameters for
$w>-1$.
A way to distinguish the two may be to analyze observational data, using
eq.~(\ref{EOS_phenexp}) rather than its linear approximation.

\medskip
Another important prediction of NNDE models lies within its initial formulation.
There is no natural way for neutrino to induce DE,
without heavier leptons inducing it as well.
We have eliminated the contribution of heavier leptons by constructing a model
where only relativistic particles have significant contributions.
However, this means that in the early universe, heavier leptons did contribute
(primordial DE).
The further we go back in time, the more contributions we have.
In order for us to discuss the cosmic history in the presence of NNDE,
we must choose a specific model, that is to determine a specific function $U(\xi,\chi)$.
As a toy model, let us use 
\begin{equation}
	\displaystyle{U=Cm^4\exp(-\xi),}
	\label{toymod}
\end{equation}
where, C is a constant.
I emphasize that the toy model is used merely for exact calculations
and all the conclusions below are model independent.

Since, unlike neutrinos, many fermions stay coupled to the cosmic plasma
in the era where $T\sim m$ (which is the era where the DE of the
specific particle dominates the expansion), the exact contribution of those
fermions is slightly different.
Figures~\ref{expxi} and~\ref{expchi} show $\exp(-\xi)$ and $\exp(-\chi)$ as a function
of $m/T$ for coupled and decoupled fermions.
It is visible that the general behavior of the model for coupled fermions,
remains the same.

\DOUBLEFIGURE[t]
{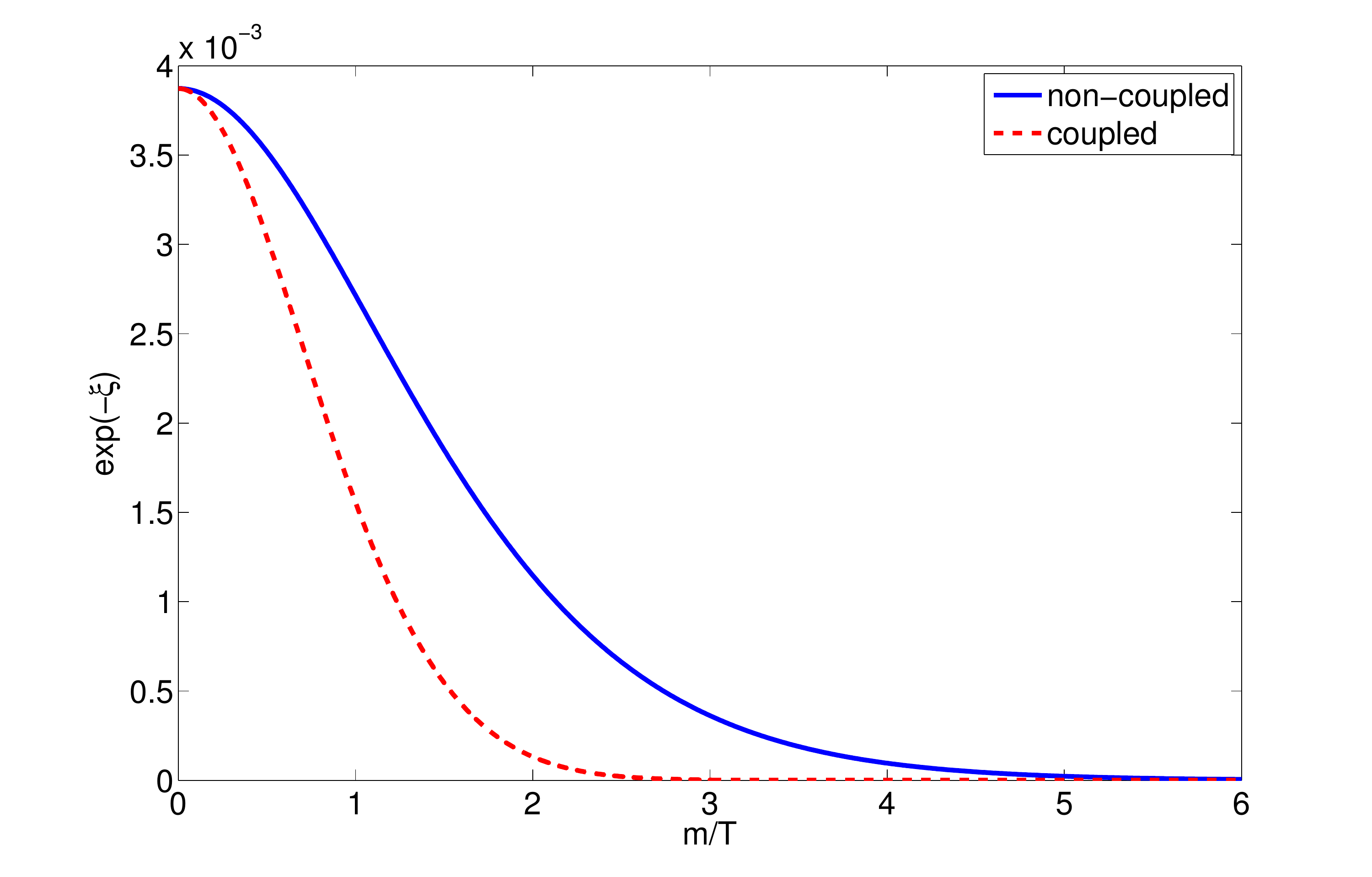, width=.52\textwidth}
{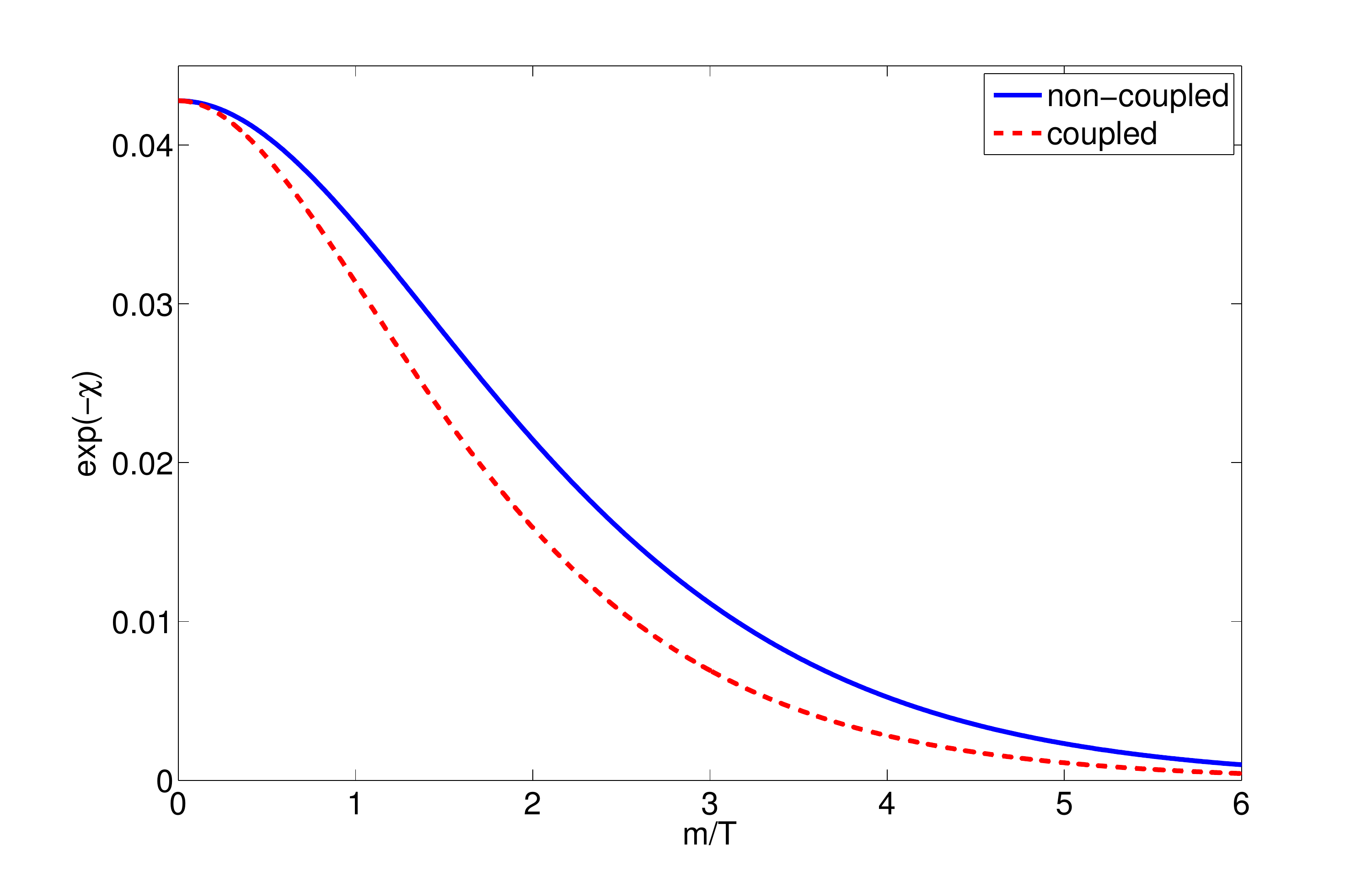, width=.52\textwidth}
{$\exp(-\xi)$ vs $m/T$ for couples and decoupled fermions. \label{expxi}}
{$\exp(-\chi)$ vs $m/T$ for couples and decoupled fermions. \label{expchi}}

Figure~\ref{Hubble} shows the expansion history, in terms of
the Hubble plot, for $\Lambda$CDM vs NNDE.
On the Hubble plot, the mass hierarchy of the different leptons is clearly visible.
The "staircase" form of the Hubble plot is in correlation with the masses of the
different leptons\footnote{
The neutrino masses are unknown so this is a schematic plot.
}.
The first DE "step" (the one closest to the present epoch) is due to
the $\mu$ neutrino, the next are due to the $\tau$ neutrino, then the electron
and so on.

\FIGURE{\epsfig{file=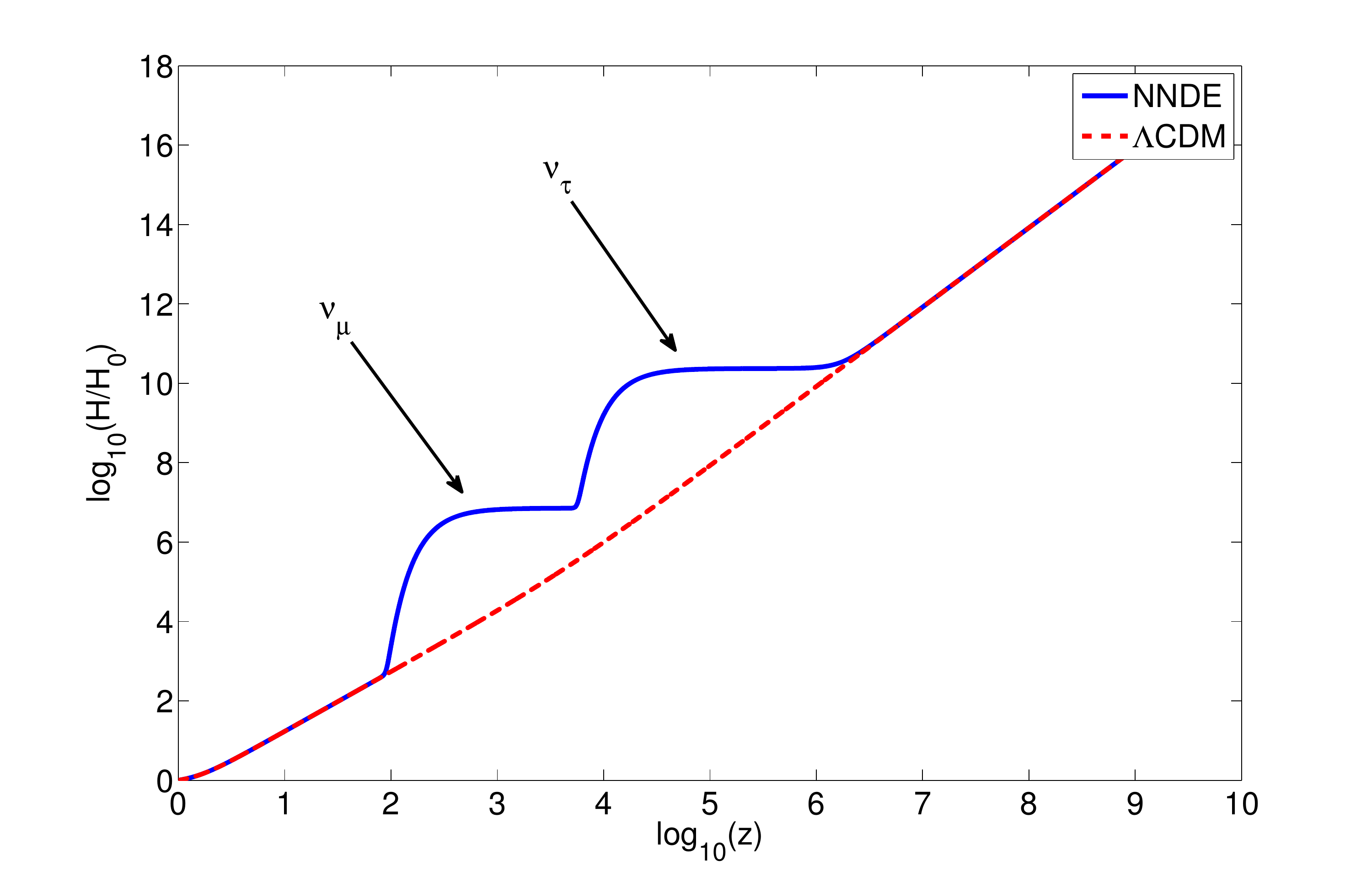,width=15cm} 
        \caption{The Hubble plot for NNDE and $\Lambda$CDM models.}
	\label{Hubble}}

\section{Constraining NNDE Using Big Bang Nucleosynthesis}

\subsection{NNDE Constraints}

Other than using the relation dictated by Eq.~(\ref{EOS_phenexp_rel}),
It is difficult (although not impossible) to constrain the NNDE model via
usual astrophysical observations, including CMB.
The reason is, that NNDE phenomenology does not differ much
from other DE models, including the cosmological constant.
The major difference lies in the primordial DE.
However, since we do not know the neutrino masses, it is difficult to state,
at which phase in the 'recent' expansion of the universe, do primordial DE
effects show themselves.

In fact, a well posed constraint on the NNDE comes from big bang
nucleosynthesis (BBN).
Since the BBN freeze-out occurs at a temperature of about $\sim1MeV$ \cite{BBN_theo},
it should be strongly affected by the electron-induced DE.
Unlike the neutrino masses, we cannot maneuver our way around this
issue.
Both the BBN ratios \cite{BBN_exp} and electron mass are measured with good precision
and a significant variation of either is out of the question.

There are two ways to ensure that the presence of the electron would not
have significant effects on the BBN, without abandoning the principles of
NNDE:

\begin{itemize}
	\item	The first way is to construct a model, where the contribution of
	any particle to DE decays before it becomes non relativistic
	(To be referred as type I NNDE).
	This would assure that by the time of the BBN freeze-out, the
	electron-induced DE is negligible.
	\item Finally, it is possible to construct a model such that the contribution
	of any particle to DE, is suppressed by a factor of several orders of
	magnitude, so it is still proportional to $m^4$, but lower
	(To be referred as type II NNDE).
	This can render the electron-induced DE low enough so
	it would not significantly affect the BBN.
\end{itemize}

To proceed, we modify the toy model from Eq.~(\ref{toymod}) to
\begin{equation}
	\displaystyle{U=Cm^4\exp(-g\xi).}
	\label{toymod2}
\end{equation}
Here, $g$ is added as a parameter which tunes the energy scale, beyond
which a particle no longer contributes to DE.

\subsection{Type I NNDE}

In this model, to assure that the electron-induced DE does not affect the BBN, we
require $g$ to be large enough, so that $\exp(-gEl)$ quickly decays for temperatures
$T<3MeV$.
Using Eq.~(\ref{EOS_exp}) obtained for the EOS, the above requirement can
be written as
\begin{equation}
	\displaystyle{g\gg\frac{1}{\beta_{\xi}}
	\left(\frac{3MeV}{m_{\rm e}}\right)^2\sim100.}
	\label{g_con_an1}
\end{equation}
At the same time, we require $1+w_0\ll1$. Thus, the neutrino mass must be small
enough, so that
\begin{equation}
	\displaystyle{m_{\nu}\ll\frac{T_{\nu}}{\sqrt{g\beta_{\xi}}}\simeq3\cdot10^{-5}eV,}
	\label{m_con_an1}
\end{equation}
where $T_{\nu}=T_{\rm CMB}(4/11)^{1/3}$.
Figure~\ref{typeI}, presents the constraints on the neutrino mass and the
parameter $g$, obtained from BBN freeze-out (neutron to proton ratio)
and the DE EOS.
The BBN freeze-out ratio, was calculated numerically using a generalization of
the model presented in \cite{BBN_theo},
\begin{eqnarray}
\displaystyle{X_{\rm n}}&=&\displaystyle{\int^{\infty}_{0}
\frac{dy}{2y^2\left(1+\cosh(1/y)\right)}\times}
\nonumber\\&&
\displaystyle{\times\exp\left(-\lambda\int^{y}_{0}dy^{\prime}H^{-1}y^{\prime2}
\left(y^{\prime}+0.25\right)^2\left(1+\exp(-1/y^{\prime})\right)\right).}
\label{BBN_eqn}
\end{eqnarray}
Here, $H$ is the Hubble constant, $\lambda\simeq3.26sec^{-1}$ and $y=T/Q$,
where, $Q\simeq1.29MeV$ is the difference between neutron and proton mass.
This model gives a good approximation of the freeze-out ratio, but more
importantly, of its sensitivity to the cosmic expansion.
To cancel out the overall error of this model, only the difference
$X_{\rm n}(\rm NNDE)-X_{\rm n}(\Lambda CDM)$ is used.

\FIGURE{\epsfig{file=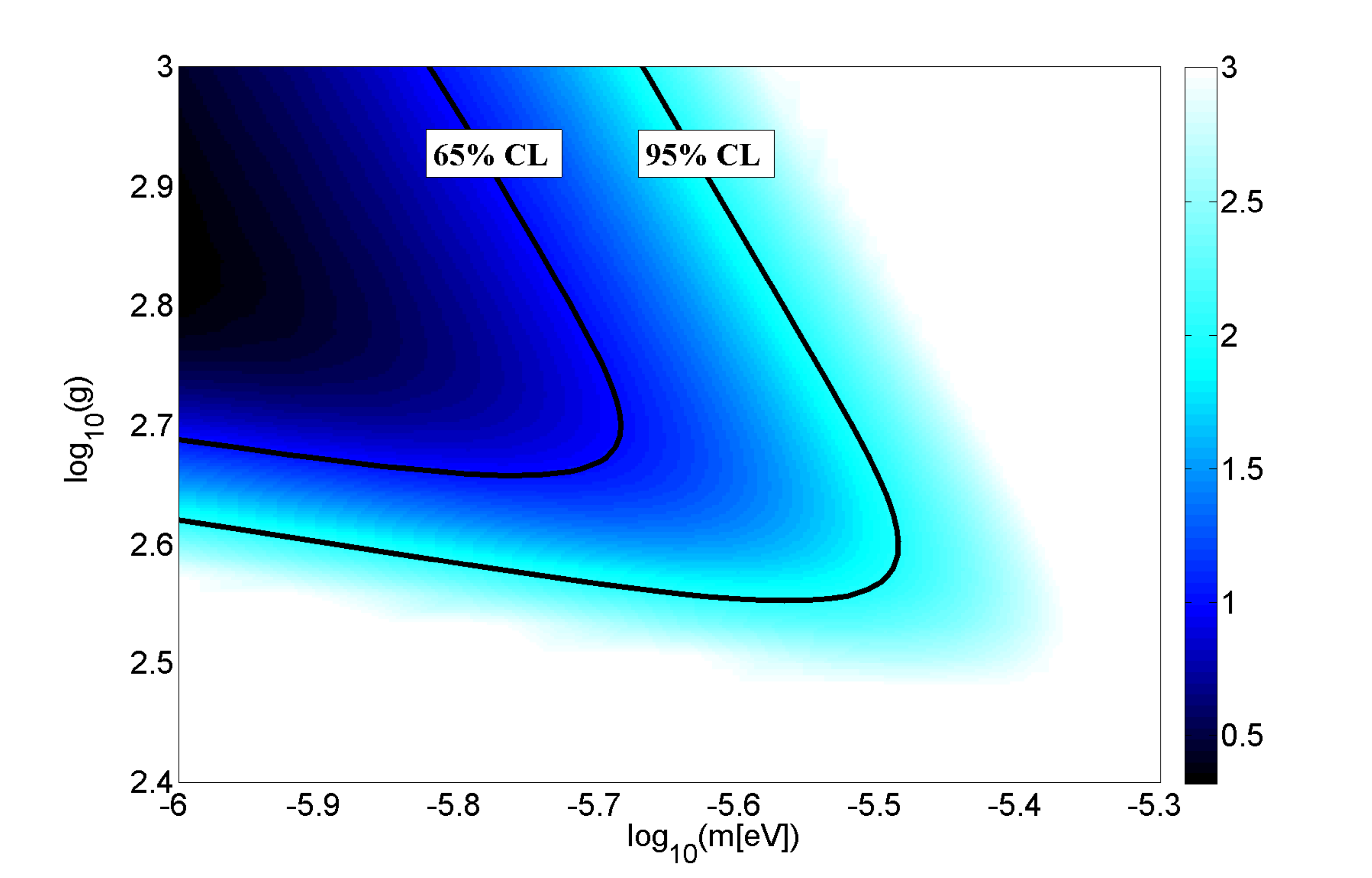,width=15cm} 
        \caption{type I NNDE constraint on $g$ and the neutrino mass from
        the DE EOS and BBN measurements.
        The color-bar signifies the sigma confidence level.}
	\label{typeI}}

\FIGURE{\epsfig{file=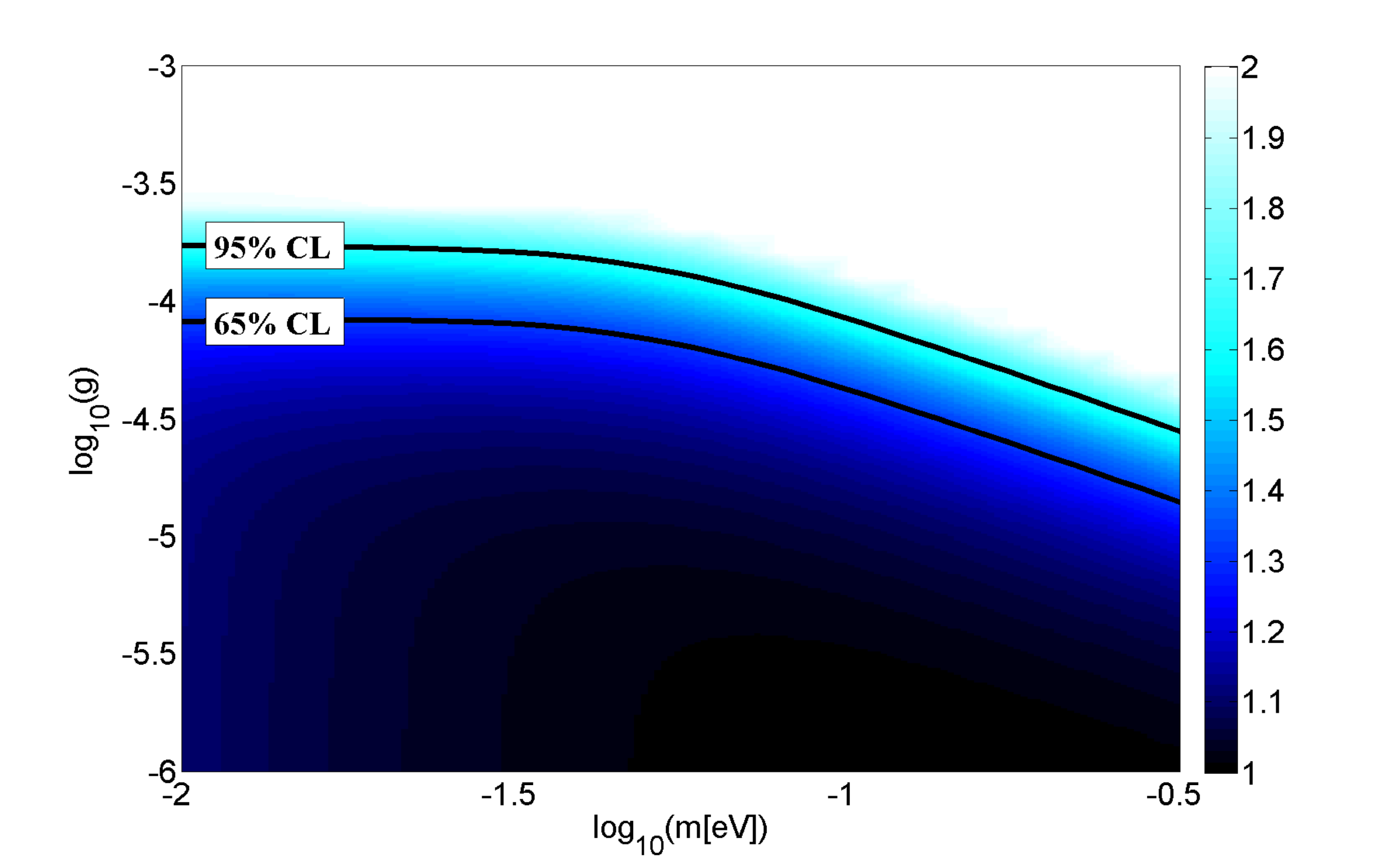,width=15cm} 
        \caption{type II NNDE constraint on $g$ and the lightest neutrino mass
        from the DE EOS and BBN measurements.
        The color-bar signifies the sigma confidence level.
        In this case, there is no constraint on the mass of the lightest neutrino.}
	\label{typeII}}
 
The 95\% confidence level constraints of this model\footnote{
The constraint on $g$ is irrelevant, since it is a model related parameter.
Using a different model, would have resulted in different constraints on its parameters,
however, {\it it would not have a significant effect on the neutrino mass constraint}.
},
are therefore
(see Figure~\ref{typeI}),
\begin{eqnarray}
\displaystyle{m_{\nu}}&\lesssim&\displaystyle{3\cdot10^{-6}eV}\,,
\nonumber\\
\displaystyle{g}&\gtrsim&\displaystyle{350}\,.
\label{mg_con_num1}
\end{eqnarray}
This is of the same order of magnitude as the value obtained in \cite{NDE2}
It is worth noting that according to these constraints on the lightest neutrino
mass and using the best estimate we have for the neutrino square mass
difference (obtained from neutrino oscillations \cite{RPP}), the muon neutrino mass
is $m_{\nu\mu}\simeq8.7\cdot10^{-3}eV$.
Therefore, in accordance with the previously noted fact, that a particle does
not contribute to DE, once the universe cools beyond a certain temperature
$T\propto m$, and knowing that the lightest neutrino dominates comic expansion
at red-shifts $z\sim1$ the muon-neutrino-induced DE, dominated
the expansion of the universe, up to red-shifts of
\begin{equation}
	\displaystyle{z_{\nu\mu} \gtrsim\frac{m_{\nu\mu}}{m_{\nu \rm e}}
	\simeq3000\sim z_{\rm eq}.}
	\label{muonneut_con}
\end{equation}
Here $z_{\rm eq}$ is the red-shift of the radiation-matter equality.
This is a remarkable result, because it tells us that the primordial DE
would not have significantly affected structure growth, and yet leaves room
for subtle affects, that will perhaps be measured in the future.

\subsection{Type II NNDE}

In this case, we require the electron-induced DE to be subdominant
(compared to the radiation energy density) all the way up to the end of the
freeze-out era $T^{\ast}\sim200keV$, so
\begin{equation}
	\displaystyle{Cm^{4}_{\rm e}\ll\kappa a_{\rm rad}T^{\ast4},}
	\label{c_con_an2}
\end{equation}
with $a_{\rm rad}$ being the radiation density constant and
$\kappa\simeq7.25$ is the number of radiative degrees of freedom,
at the corresponding temperature.
Since, today we have $Cm^{4}_{\nu}=\rho_{\rm DE}$, eq.~\ref{c_con_an2}
implies, for the neutrino mass,
\begin{equation}
	\displaystyle{m_{\nu}>\frac{m_{\rm e}}{T^{\ast}}
	\left(\frac{\rho_{\rm DE}}{\kappa a_{\rm rad}}\right)^{1/4}
	\simeq4\cdot10^{-3}eV.}
	\label{m_con_an2}
\end{equation}
For this mass range, we can no longer assume only the lightest neutrino
contributes.
Therefore, for this model, the calculations include all three neutrino generations.
The relations between the different neutrino masses, were taken to be the
nominal values from \cite{RPP},
\begin{eqnarray}
\displaystyle{\Delta m^{2}_{\mu \rm e}}&=&\displaystyle{7.6\cdot10^{-5}eV^2}\,,
\nonumber\\
\displaystyle{\Delta m^{2}_{\tau \mu}}&=&\displaystyle{2.4\cdot10^{-3}eV^2}\,.
\label{neut_mix_mass}
\end{eqnarray}
Figure~\ref{typeII}, presents the constraints of this model on the lightest neutrino
mass and the parameter $g$.
There are no constraints on the neutrino mass for this model.
However, we see that the maximal allowed value of $g$ is inversely proportional
to $m_{\nu\tau}$.
In this case, all three neutrino species, are involved in generating todays DE.
Therefore, there are no available particles to induce primordial DE in proximity
to the radiation-matter equality, and
thus, no observable effects on the cosmic structure formation.

\section{Summary and Discussion}

A natural model for neutrino dark energy (NNDE), free of exotic particles or couplings,
was presented.
Basing the theory only on naturalness and the assumption that todays DE is driven by
the neutrino, several unique predictions were derived.
A specific relation $w_{\rm a}(w_0)$ governing the linear evolution approximation of
the EOS, follows from NNDE.
Verifying the above relation, can strongly support this model.
Although, quintessence thawing models can produce similar $w_{\rm a}(w_0)$ dependance,
distinguishing between the two may be possible via a more advanced analysis
of observations, using second order expansion of the EOS.

NNDE also requires the existence of primordial DE, driven by heavier
particles, dominating the expansion of the early universe.
To prevent the electron-induced primordial DE from affecting the BBN,
two methods were proposed.
The first, entitled type I NNDE, is to assure that the electron-induced
DE decays prior to the BBN freeze-out phase.
The second method, entitled type II NNDE, is to assure that the magnitude
of the electron-induced DE, is small enough to have little affect of the BBN.
Using the above methods, constrains on NNDE are placed.

Type I NNDE, provides new constraints on the mass of the lightest
neutrino, requiring $m_{\nu}<3\cdot10^{-6}eV$.
In addition, type I NNDE, predicts relic primordial DE, induced by the muon
neutrino, in proximity to the radiation-matter equality phase.
This would mildly affect structure formation and the CMB.
Observing these effects could prove as the main finger-print of the NNDE.

\acknowledgments

I would like to thank Aharon Davidson and Ram Brustein for insightful
discussions and many useful comments.


\begin{thebibliography}{999}


\bibitem{DE_discovery} 
A. G. Riess et al. {\it Observational evidence from supernovae for an accelerating universe and a cosmological constant}, \asj{116}{1998}{1009};
S. Perlmutter et al. {\it Measurements of Omega and Lambda from 42 high redshift supernovae}, \asj{517}{1999}{565}.

\bibitem{WMAP} 
E. Komatsu et al. {\it Seven-Year Wilkinson Microwave Anisotropy Probe (WMAP) Observations: Cosmological Interpretation}, \arXivid{1001.4538}.

\bibitem{SDSS} 
U. Seljak, {\it Cosmological parameter analysis including SDSS Ly-alpha forest and galaxy bias: Constraints on the primordial spectrum of fluctuations, neutrino mass, and dark energy}, \prd{71}{2005}{103515};
W. J. Percival et al. {\it Baryon Acoustic Oscillations in the Sloan Digital Sky Survey Data Release 7 Galaxy Sample}, \newjournal{Mon.\ Not.\ Roy.\ Astron.\ Soc.\ }{MNRAA}{401}{2010}{2148}

\bibitem{Hubble_DEC} 
A. G. Riess et al. {\it A Redetermination of the Hubble Constant with the Hubble Space Telescope from a Differential Distance Ladder}, \apj{699}{2009}{539}.


\bibitem{DGP} 
G. R. Dvali, G. Gabadadze and M. Porrati, {\it 4-D gravity on a brane in 5-D Minkowski space}, \plb{485}{2000}{208};
C. Deffayet, {\it Cosmology on a brane in Minkowski bulk}, \plb{502}{2001}{199};
C. Deffayet, G. R. Dvali and G. Gabadadze, {\it Accelerated universe from gravity leaking to extra dimensions}, \prd{65}{2002}{044023}.

\bibitem{ModGrav} 
A. D. Dolgov and M. Kawasaki, {\it Can modified gravity explain accelerated cosmic expansion?}, \plb{573}{2003}{1};
S. M. Carroll, V. Duvvuri, M. Trodden and M. S. Turner, {\it Is cosmic speed - up due to new gravitational physics?}, \prd{70}{2004}{043528};
S. M. Carroll et al. {\it The Cosmology of generalized modified gravity models}, \prd{71}{2005}{063513};
E. J. Copeland, M. Sami and S. Tsujikawa, {\it Dynamics of dark energy}, \ijmpd{15}{2006}{1753};
W. Hu and I. Sawicki, {\it Models of f(R) Cosmic Acceleration that Evade Solar-System Tests}, \prd{76}{2007}{064004};
T. P. Sotiriou and V. Faraoni, {\it f(R) Theories Of Gravity}, \rmp{82}{2010}{451}.

\bibitem{Quintessence} 
R. R. Caldwell, R. Dave and P. J. Steinhardt, {\it Cosmological imprint of an energy component with general equation of state}, \prl{80}{1998}{1582};
I. Zlatev, L.-M. Wang, P. J. Steinhardt, {\it Quintessence, cosmic coincidence, and the cosmological constant}, \prl{82}{1999}{896};
L. Amendola, {\it Coupled quintessence}, \prd{62}{2000}{043511};
R. R. Caldwell and M. Kamionkowski, {\it The Physics of Cosmic Acceleration}, \arnps{59}{2009}{397}.

\bibitem{Chaplygin} 
N. Bilic, G. B. Tupper and R. D. Viollier, {\it Unification of dark matter and dark energy: The Inhomogeneous Chaplygin gas}, \plb{535}{2002}{17};
M.C. Bento, O. Bertolami and A. A. Sen, {\it Generalized Chaplygin gas, accelerated expansion and dark energy matter unification}, \prd{66}{2002}{043507}.


\bibitem{DEC} 
W. M. Wood-Vasey et al. {\it Observational Constraints on the Nature of the Dark Energy: First Cosmological Results from the ESSENCE Supernova Survey}, \apj{666}{2007}{694};
E. L. Wright, {\it Constraints on Dark Energy from Supernovae, Gamma Ray Bursts, Acoustic Oscillations, Nucleosynthesis and Large Scale Structure and the Hubble constant}, \apj{664}{2007}{633};
Y. Wang and P. Mukherjee, {\it Observational Constraints on Dark Energy and Cosmic Curvature}, \prd{76}{2007}{103533};
S. W. Allen et al. {\it Improved constraints on dark energy from Chandra X-ray observations of the largest relaxed galaxy clusters}, \newjournal{Mon.\ Not.\ Roy.\ Astron.\ Soc.\ }{MNRAA}{383}{2008}{879};
S. Bhattacharya and A. Kosowsky, {\it Dark Energy Constraints from Galaxy Cluster Peculiar Velocities}, \prd{77}{2008}{083004};
Y. Wang, {\it Model-Independent Distance Measurements from Gamma-Ray Bursts and Constraints on Dark Energy}, \prd{78}{2008}{123532};
R. Tsutsui et al. {\it Constraints on $w_0$ and $w_a$ of Dark Energy from High Redshift Gamma Ray Bursts}, \newjournal{Mon.\ Not.\ Roy.\ Astron.\ Soc.\ }{MNRAA}{394}{2009}{L31};
R. de Putter, O. Zahn and E. V. Linder, {\it CMB Lensing Constraints on Neutrinos and Dark Energy}, \prd{79}{2009}{065033};
M. Hicken et al. {\it Improved Dark Energy Constraints from ~100 New CfA Supernova Type Ia Light Curves}, \apj{700}{2009}{1097};
C.-W. Chen, P. Chen and J.-A. Gu, {\it Constraints on the Phase Plane of the Dark Energy Equation of State}, \plb{682}{2009}{267};
Y. Wang, {\it Distance Measurements from Supernovae and Dark Energy Constraints}, \prd{80}{2009}{123525};
H. Li and J.-Q. Xia, {\it Constraints on Dark Energy Parameters from Correlations of CMB with LSS}, \newjournal{J.\ Cosmo.\ Astropart.\ Phys.\ }{JCAP}{04}{2010}{026};
N. Pan, Y. Gong and Z.-H. Zhu, {\it Improved cosmological constraints on the curvature and equation of state of dark energy}, \cqg{27}{2010}{155015};
Q.-J. Zhang and Y.-L. Wu, {\it Time-Varying Dark Energy Constraints From the Latest SN Ia, BAO and SGL}, \arXivid{1008.0930}.


\bibitem{RPP} 
C. Amsler et al. {\it Review of Particle Physics}, \plb{667}{2008}{1}.


\bibitem{CNB} 
C. Quigg, {\it Cosmic Neutrinos}, \arXivid{0802.0013};
R. J. Michney and R. R. Caldwell, {\it Anisotropy of the Cosmic Neutrino Background},  \newjournal{J.\ Cosmo.\ Astropart.\ Phys.\ }{JCAP}{01}{2007}{014}.


\bibitem{NeutMix_DE} 
M. Blasone et al. {\it Neutrino mixing contribution to the cosmological constant}, \pla{323}{2004}{182};
A. Capolupo, S. Capozziello and G. Vitiello, {\it Neutrino mixing as a source of dark energy}, \pla{363}{2007}{53};
A. Capolupo, S. Capozziello and G. Vitiello, {\it Dark energy and particle mixing}, \pla{373}{2007}{601}.

\bibitem{CLEP_DE} 
E. I. Guendelman and A. B. Kaganovich, {\it Neutrino dark energy}, \hepth{0411188}.

\bibitem{GoldstoneB_DE} 
P.-H. Gu, H.-J. He and U. Sarkar, {\it Dirac neutrinos, dark energy and baryon asymmetry}, \newjournal{J.\ Cosmo.\ Astropart.\ Phys.\ }{JCAP}{11}{2007}{016}.

\bibitem{Vacuum_DE} 
H. J. de Vega and N. G. Sanchez, {\it Dark energy is the cosmological quantum vacuum energy of light particles. The axion and the lightest neutrino}, \astroph{0701212}.

\bibitem{NDE1} 
T. Goldman, G. J. Stephenson Jr., P. M. Alsing and B. H. J. McKellar, {\it A Possible Connection Between Massive Fermions and Dark Energy}, \arXivid{0905.4308}.

\bibitem{NDE2} 
G. Lambiase, H. Mishra and S. Mohanty, {\it Dark energy from Neutrinos and Standard Model Higgs potential}, \arXivid{1006.4461}.

\bibitem{MaVaNs_basic} 
R. Fardon, A. E. Nelson and N. Weiner, {\it Dark energy from mass varying neutrinos}, \newjournal{J.\ Cosmo.\ Astropart.\ Phys.\ }{JCAP}{10}{2004}{005};
R. D. Peccei, {\it Neutrino models of dark energy}, \prd{71}{2005}{023527}.

\bibitem{MaVaNs} 
R. Fardon, A. E. Nelson and N. Weiner, {\it Supersymmetric theories of neutrino dark energy}, \jhep{03}{2006}{042};
R. Horvat, {\it Mass-varying neutrinos from a variable cosmological constant},  \newjournal{J.\ Cosmo.\ Astropart.\ Phys.\ }{JCAP}{01}{2006}{015};
S. Antusch, S. Das and K. Dutta, {\it Phenomenology of Hybrid Scenarios of Neutrino Dark Energy}, \newjournal{J.\ Cosmo.\ Astropart.\ Phys.\ }{JCAP}{10}{2008}{016};
O. E. Bjaelde and S. Hannestad, {\it Neutrino Dark Energy With More Than One Neutrino Species}, \prd{81}{2010}{063001}.


\bibitem{BBN_theo} 
V. F. Mukhanov, {\it Nucleosynthesis without a computer}, \ijtp{43}{2004}{669}.

\bibitem{BBN_exp} 
R. H. Cyburt, B. D. Fields and K. A. Olive, {\it Primordial nucleosynthesis in light of WMAP},  \plb{567}{2003}{227};
A. Coc et al. {\it Updated Big Bang nucleosynthesis confronted to WMAP observations and to the abundance of light elements}, \apj{600}{2004}{544};
K. A. Olive and E. D. Skillman, {\it A Realistic determination of the error on the primordial helium abundance: Steps toward non-parametric nebular helium abundances}, \apj{617}{2004}{29}.

\end{thebibliography}
\end{document}